# *Cognition coming about: self-organisation and free-energy*

a Commentary on Wright, J. J., & Bourke, P. D. (2020). The growth of cognition: free energy minimization and the embryogenesis of cortical computation. *Physics of Life Reviews*.

**Authors**
Inês Hipólito 1,2
Maxwell Ramstead 3,4,5
Axel Constant 6
Karl J. Friston 5

1. Faculty of Arts, Social Sciences, and Humanities, University of Wollongong, Australia.
2. Institute of Psychiatry, Psychology and Neuroscience (IoPPN), King's College London, United Kingdom.
3. Division of Social and Transcultural Psychiatry, Department of Psychiatry, McGill University, Canada.
4. Culture, Mind, and Brain Program, McGill University, Canada.
5. Wellcome Centre for Human Neuroimaging, University College London, United Kingdom.
6. Charles Perkins Centre, The University of Sydney, Australia.

**Abstract**

Wright and Bourke's compelling article rightly points out that existing models of embryogenesis fail to explain the mechanisms and functional significance of the dynamic connections among neurons. We pursue their account of Dynamic Logic by appealing to the Markov blanket formalism that underwrites the Free Energy Principle. We submit that this allows one to model embryogenesis as self-organisation in a dynamical system that minimises free-energy. The ensuing formalism may be extended to also explain the autonomous emergence of cognition, specifically in the brain, as a dynamic self-assembling process.

Wright and Bourke's [1] compelling treatment casts embryogenesis as the formation—in living tissue—of a network endowed with a low degree of separation, where initially weak cell-to-cell connectivity is strengthened, eventually converging to a small world architecture. The authors rightly note that existing models of embryogenesis fail to explain the mechanisms and functional significance of the dynamic connections entailed by individual neurons associating to their preferences. They propose a framework to model cortical dynamics in embryogenesis that draws on the Free-Energy Principle (FEP) and Dynamic Logic (DL). Wright and Bourke [1] contend that the FEP enables us to explain the elementary laws of brain organisation and development, and that DL allows us to specify the information-processing mechanisms operative in the brain. This commentary attempts to mitigate some of the authors' concerns as to what constitutes an appropriate explanation of neuronal assembly formation. It aims to do so by complementing DL with the Markov blanket formalism that underwrites the FEP. Markov blankets extend Wright and Bourke's model of embryogenesis, by allowing us to model embryogenesis as the self-organisation of boundaries in a dynamical system that minimises free-energy. This formalism further explains the autonomous emergence of cognition, specifically in the brain, as a dynamic self-assembling process.

A Markov blanket is a set of states (so called blanket states) that separates the internal or systemic states of interest from external states that are not part of the system [2,3]. Given the blanket states, internal and external states are conditionally independent—as in they can only affect one another vicariously through blanket states. At the same time, they are coupled by the exchanges with blanket states. The latter are generally partitioned into sensory and active states; where sensory states influence but are not influenced by internal states, and active states influence but are not influenced by external states.

Studies of neural dynamics using dynamic causal modelling [4,5] provide compelling evidence that synaptic connections conform to the conditional independence structure of a Markov blanket [3,6,7]. This warrants a partition of the states of a system (e.g., the brain) into internal, blanket, and external states. These states map onto known neurophysiology at several levels of organization. For instance, on the level of individual neurons, the internal states (e.g., ion channel conductance) of one neuron are conditionally independent from the internal states of other neurons (i.e., the external states the first). They interact, however, through presynaptic and postsynaptic voltages (sensory and active states, respectively). In a nutshell, this means that internal states evolve based upon internal and blanket states (sensory and active states), but not external states – given the conditional independency.

In the specific context of cortical dynamics in embryogenesis, the presence of a Markov blanket is useful to the extent that it holds explanatory value with respect to connectivity. The influences of blanket states (e.g., membrane potentials) vicariously enable internal and external states (ion channel conductance) to connect with each other (see Fig 1.).

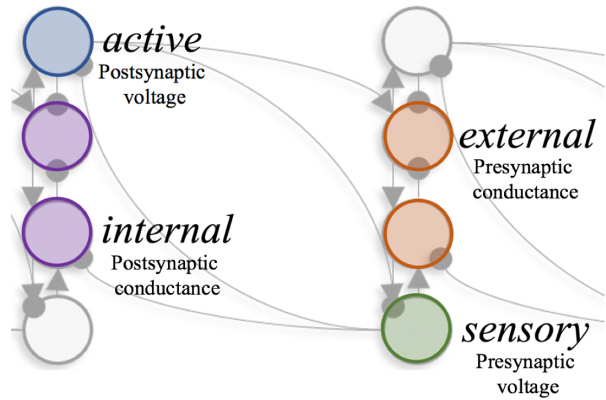

**Figure 1.** Illustration of a postsynaptic neuron (on the left) and of a presynaptic neuron (on the right). A membrane potential (active state) of a neuron can only be directly influenced by the conductance of a neuron (internal states). Conversely, the presynaptic potentials (sensory states) of other neurons influence the internal states (adapted from Hipólito et al. 2020).

More specifically, the dependencies established by a Markov blanket imply a circular causality, where external states, i.e., the presynaptic conductance, cause changes in internal states, i.e., the postsynaptic conductance, through the sensory states, i.e. presynaptic voltage. The internal states, in turn, couple back to external states via active states, i.e. postsynaptic voltage.

It is important to note that Markovian partitioning does not necessarily correspond to physical boundaries, for example of neurons. As argued by Hipólito and colleagues [3], the epistemological value of Markov blankets is that they can be applied to abstract states—not only to physical boundaries—and hence, they can be used to highlight the probabilistic relations or couplings between areas, regions, or networks. The epistemological value of this formalism is precisely that the same formalism applies even in the absence of clear spatial boundaries.

At the scale above, the Markovian asymmetry can be recapitulated in a canonical microcircuit model, which allows us to represent slower and larger units of organization. On macroscales, the networks themselves become the active, sensory, internal, and external states [8–10]. Thus we can depict cortical columns and brain regions, in turn, as active, sensory, internal, and external states, composed of other Markovian states (for details see [3,7]). Bounded assemblies at all scales (from single neurons, through to brain regions and brain-wide networks) form spontaneously. This is consistent with the self-organisation of complex systems defined as structures that maintain their integrity under changing conditions.

With this formulation in place, we arrive at the epistemological value held by the Markovian formalisms of cortical dynamics in embryogenesis. We can investigate how the minimisation of free-energy ensures that the brain optimises the neuronal assembly in terms of entropy and surprise reduction (aka active inference [11–13]). Surprise is the self-information under the solution to the Fokker-Plank equation that yields nonequilibrium steady state. These patterns ensure that the agent keeps surprise (and entropy) at bay, in the face of environmental pressures; pressures that favoured the selection of those patterns in the first-place relative to available developmental inputs [14–2]. What the agent inherits, then, are genetic and epigenetic predispositions to enact different trajectories in the space of active and internal states [3] by encountering expected developmental inputs [15].

In conclusion, the Markov blanket formalisms of the FEP complements the use of FEP and DL in the target article, by allowing us to model embryogenesis as the self-organisation of boundaries in a dynamical system that minimises free-energy. During phylogenetic development, the embodied brain requires interaction with the environment to maintain its organised patterns. Cortical dynamics in embryogenesis, we submit, can be defined by appealing to the Markov blanket formalism at each level—from single neurons, through to brain regions and brain-wide networks—which is the conceptual basis of effective connectivity under dynamical causal modelling [16]. Neuronal self-assembly is then specified epigenetically by, and fulfilled behaviourally and developmentally under, active inference.

## References

1. Wright JJ, Bourke PD. The growth of cognition: Free energy minimization and the embryogenesis of cortical computation. Phys Life Rev. 2020. doi:10.1016/j.plrev.2020.05.004

2. Ramstead MJD, Badcock PB, Friston KJ. Answering Schrödinger's question: A free-energy formulation. Phys Life Rev. 2018;24: 1–16.

3. Hipólito I, Ramstead M, Convertino L, Bhat A, Friston K, Parr T. Markov Blankets in the Brain. arXiv [q-bio.NC]. 2020. Available: http://arxiv.org/abs/2006.02741

4. Bastos AM, Usrey WM, Adams RA, Mangun GR, Fries P, Friston KJ. Canonical microcircuits for predictive coding. Neuron. 2012;76: 695–711.

5. Moran R, Pinotsis DA, Friston K. Neural masses and fields in dynamic causal modeling. Front Comput Neurosci. 2013;7: 57.

6. Parr T, Da Costa L, Friston K. Markov blankets, information geometry and stochastic thermodynamics. Philos Trans A Math Phys Eng Sci. 2020;378: 20190159.

7. Friston KJ, Fagerholm ED, Zarghami TS, Parr T, Hipólito I, Magrou L, et al. Parcels and particles: Markov blankets in the brain. arXiv [q-bio.NC]. 2020. Available: http://arxiv.org/abs/2007.09704

8. Razi A, Kahan J, Rees G, Friston KJ. Construct validation of a DCM for resting state fMRI. Neuroimage. 2015;106: 1–14.

9. Sharaev MG, Zavyalova VV, Ushakov VL, Kartashov SI, Velichkovsky BM. Effective Connectivity within the Default Mode Network: Dynamic Causal Modeling of Resting-State fMRI Data. Front Hum Neurosci. 2016;10: 14.

10. Betzel RF, Byrge L, He Y, Goñi J, Zuo X-N, Sporns O. Changes in structural and functional connectivity among resting-state networks across the human lifespan. NeuroImage. 2014. pp. 345–357. doi:10.1016/j.neuroimage.2014.07.067

11. Friston KJ, FitzGerald T, Rigoli F, Schwartenbeck P, Pezzulo G. Active inference: a process theory. Neural Comput. 2016;29: 1–49.

12. Friston KJ. A free energy principle for a particular physics. arXiv [q-bio.NC]. 2019. Available: http://arxiv.org/abs/1906.10184

13. Ramstead MJD, Kirchhoff MD, Friston KJ. A tale of two densities: active inference is enactive inference. Adapt Behav. 2019; 1059712319862774.

14. Oyama S, Griffiths PE, Gray RD. Cycles of Contingency: Developmental Systems and Evolution. MIT Press; 2003.

15. Constant A, Clark A, Kirchhoff M, Friston KJ. Extended active inference: constructing predictive cognition beyond skulls. Mind and Language. 2019. Available: http://sro.sussex.ac.uk/id/eprint/88369/

16. Zarghami TS, Friston KJ. Dynamic effective connectivity. Neuroimage. 2020;207: 116453.